# Space storm measurements of 17 and 21 April 2002 Forbush effects from ARTEMIS-IV coronagraph, Athens NEUTRON MONITOR station and CORONAS-F satellite

C. Caroubalos[1], H. Mavromichalaki[2], A. Hillaris[2], X. Moussas[2], P. Preka-Papadema[2], I. Polygiannakis[2,*], C. Sarlanis[2], G. Souvatzoglou[2], M. Gerontidou[2], C. Plainaki[2], S. Tatsis[2], S. N. Kuznetsov[3], I. N. Myagkova[3] and K. Kudela[4]

[1]{Department of Informatics, University of Athens, Panepistimioupolis Zografos 15784, Athens Greece }

[2]{Department of Physics, University of Athens, Panepistimioupolis Zografos 15784, Athens Greece}

[3]{Scobeltsyn Institute of Nuclear Physics, Moscow State University, 119899 Moscow Russia}

[4]{Institute of Experimental Physics, SAS, 043 53 Kosice, Slovakia}

[*]{Deceased}

Correspondence to A. Hillaris (ahilaris@cc.uoa.gr)

## Abstract

In this report we present two complex eruptive solar events and the associated Cosmic Ray effects (Forbush decrease). We use combined recordings from a number of Earthbound Receivers, Space Experiments and data archives (such as the ARTEMIS-IV Radio spectrograph, the Athens NEUTRON MONITOR, the LASCO CME Lists, the SONG of the {CORONAS-F} satellite, etc.). The influence of solar transients on the interplanetary medium conditions and the cosmic ray flux is analysed and discussed. The observed time sequence of events of this time period indicates that the initiation of CMEs is closely related to the appearance of type II and IV radio bursts and strong solar flares. Their effects extend from the lower corona to the near Earth vicinity affecting Cosmic Ray measurements and space weather. As regards the Forbush decrease our data indicate significant amplification at the presence of a MHD shock.

# 1 Introduction

The response of the Magnetosphere and the near Earth environment to the Solar Transients has been dubbed "*Space Weather*" (cf. for example Brekke (2001)). Geomagnetic Storms, Forbush decrease (Cane (2000)) and increased particle fluxes are some of the most important manifestations of Space Weather.

The most severe disturbances are associated with Coronal Mass Ejections MHD shocks and Flares; these are known as "*non recurrent decreases*" and are characterised by a sudden onset and a more gradual recovery; when both MHD shock and CME are present they result in a classical "*two step*" Forbush decrease (cf. Cane 2000 for a review), the steps being attributed to each of the two respectively. More often than not, the energetic solar events are preceded or accompanied by Solar Radio Bursts (mostly type II and IV); these are excited by the above mentioned disturbances, thus being their manifestation in the metric, Decametric and hectometric-kilometric range, or they are associated with the complex energy release accompanying their launch. The exact details of this connection are, as yet, uncertain. Therefore the establishment of an association between Solar Bursts and magnetospheric disturbances accurate enough for Space Weather Forecasting remains an open research issue.

In this report we present two complex solar events originating both from AR 9906 on the 17th and the 21st April 2002. We study, also, the associated space weather and cosmic ray effects on Earth

# 2 Instrumentation & Observations

We have used information from a number of instruments and on line data archives:

- The Dynamic Spectra in the range 110-698 (at a rate of 10 sec$^{-1}$) were from the ARTEMIS-IV solar radio spectrograph at Thermopylae, Greece (38 49N, 22 41E, cf Caroubalos et al 2001 for a detailed technical description of the ARTEMIS-IV Radio spectrograph); they were supplemented with spectra from the Nançay Decametric Array[1] in the 70-20 MHz range.

- The Neutron Monitor Recordings of the hadronic component of the cosmic ray intensity were from the Athens Neutron Monitor[2] (Mavromichalaki et al, 2001) with cut-off rigidity 8.53 GeV and the Oulu Finland Neutron Monitor (cut-off rigidity 0.77 GeV).

---

[1] http://www.obs-nancay.fr/html-an/quicklook

[2] http://cosray.phys.uoa.gr

- The CME data from the LASCO coronagraph (Brueckner et al, 1995) were obtained from the LASCO event list[3] (Yashiro et al , 2001).

- The proton fluxes were recorded by SONG on board the CORONAS-F satellite in the range 1-5, 14-26, 26-50 and 50-90 MeV. We have also used data from the NOAA Solar Geophysical Data catalogues[4] for the GOES SXR and particle Fluxes ($E_p > 10$ MeV) as well as the total magnetic field.

In order to associate distance from the solar centre (used in the CME Height Time Plots as y-axis) and corresponding frequency (used as y-axis in the dynamic spectra) we have assumed that the radio bursts were recorded in the first harmonic of the local plasma frequency.

$$f (\text{in } KHz) = 18\sqrt{N_e} \qquad (1)$$

We have used the Newkirk (1961) density ($N_e$, in cm$^{-3}$) vs. height (R, in $R_{SUN}$) coronal model:

$$N_e = 4.2 \cdot 10^4 \cdot 10^{4.32/(R-1)} = 4.2 \cdot 10^4 \cdot \exp\left(\frac{9.947}{R-1}\right) \qquad (2)$$

The major phases of both events are shown in the Table and in the figures.

The 17 April, 2002 Event manifests an extensive energy release associated with a CME ejection and an MHD shock. The event induces a particle flux increase while the arrival of the shock at the Earth is marked by a short but abrupt increase of the magnetic field. Both the shock and the CME produce a Forbush decrease on the neutron monitor recordings {*double Forbush decrease*}.

The 21 April 2002 Event manifests an extensive energy release associated with the ejection of two CMEs in close succession but not an Earthbound MHD shock since by April the 21st AR9906 is at the solar limb. A similar increase in particle flux and a single Forbush decrease are recorded

---

[3] http://cdaw.gsfc.nasa.gov/CME_list

[4] http://www.sel.noaa.gov

## 3  Results and Discussion

A joint campaign of the Artemis-IV Group, the Cosmic Ray Group of the Athens University and the Coronas-F team has started attempting to associate multi instrument recordings in order to connect transient solar activity with subsequent space weather effects on the Earth's magnetosphere. In this report, combined observations of solar transients including strong flares, type II and IV radio bursts and halo-CME and of the following Cosmic Ray decreases in April 2002 are presented.

It is well known that certain types of intense transient solar activity, detected by type II and IV solar bursts, result in interplanetary and geomagnetic phenomena, perturbed space weather and cosmic ray storms. In particular:

- The MHD Shocks: They excite *type II bursts* which trace their passage through the solar corona; their radio emission is due either to energetic electrons accelerated at the shock front or plasma turbulence excited by the shock, they originate either by a flare blast wave or by a CME front or flanks (Maia et al (2000); Aurass (1997)).

- The CMEs: They excite type II bursts and type IV continua: The type II bursts originate in the corresponding MHD shock, the continua (under the general name *type IV bursts*), on the other hand, represented the radiation of energetic electrons trapped within magnetic clouds, CMEs and plasmoids.

Although there is no clear and unambiguous picture about the CME-type II relationship, since the origins of the latter vary, Caroubalos et al (2004) presented statistics on the association of type II, occasionally followed by type IV bursts and CMEs from a data sample covering the period from 1998 to 2000. The association probability was found to be 70% in accordance with previous results (Sheeley et al, 1984) yet it increases from 70% to 100% in the case of complex II/IV events. Furthermore the general trend established is that the CME probability for both type II and type IV bursts accompanied by a $H_\alpha$ flare increases with brightness, area and duration or, for GOES/SXR flares, with peak intensity, total flux and duration. When both CMEs and shock effects are present the resulting cosmic ray event is called Forbush decrease.

On April 17, 2002 EIT and LASCO observed a full halo CME event (Fig. 4); it appeared on C2 starting at 08:26 UT as a large loop front with a cavity over the west hemisphere. GOES reported a long duration M2.3 SXR flare in the same time period. EIT observed a flare in AR9906 at S14W34 beginning at 07:46 UT. These events are probably different manifestations of an extended energy release. A type II burst followed by a type IV burst was recorded by the ARTEMIS IV solar radio-spectrograph from 08:03 to 10:27 UT in the frequency range from 110 to 687 MHz. The dynamic spectrum exceeded the ARTEMIS-IV frequency range extending into the decametric frequency regime covered by the Nançay Decametric Array. A composite spectrum is presented in figure 3.

An abrupt decrease of amplitude 2.5% in the cosmic ray intensity recorded by the Athens NM took place on April 17 at 10:00 UT presumably caused by a CME on 15 April, 2002. At the same time a sharp increase during this event was followed by a second decrease on April 18 at 08:00 UT. The total amplitude of this decrease was about 3% (Fig. 5). Moscow NM station (cut-off rigidity 2.30GV) recorded this event with amplitude of 5.5% and Oulu NM station (cut-off rigidity 0.77GV) with 5.8 %. This event is characteristic of a Forbush decrease of two steps (Cane 2000). The second decrease caused by the shock wave originated by the type II radio burst and the halo- CME of the April 17 at 08.26 UT, as recorded by the ARTEMIS IV of the Solar Observatory. Later, the same AR9906 at S14W84 gave the solar flare X1/1F on April 21, 2002 at 1.51 UT (Table) that corresponds to a Forbush decrease recorded by the Athens NM with an increase on April 21, 18.00 UT and a gradual depression at 08.00 UT of April 22, 2002.

## 4  Conclusions

Two composite events each consisting of a sequence of solar transients (flares, type II and IV radio bursts and CMEs), as well as their effects on the near Earth environment (Forbush decrease, and geomagnetic storms), were presented. The observations indicate that the CME lift-off is well associated with type II and IV radio bursts and SXR flares, while the Forbush decrease is mainly associated with the MHD shock, driven by the CME front, which appears as type II burst in its first stages

**Acknowledgements**

This work was financially supported by the Research Committee of the University of Athens. The LASCO CME catalogue is generated and maintained by the Centre for Solar Physics and Space Weather, The Catholic University of America in cooperation with the Naval Research Laboratory and NASA. SOHO is a project of international cooperation between ESA and NASA. The Solar Geophysical Data Catalogue is compiled and maintained by the US Department of Commerce.

Table 1. Summary of the 17 April 2002 & 21 April 2002 Events

| Event 17 April 2002 | | | Event 21 April 2002 | | |
|---|---|---|---|---|---|
| Time (UT) | Activity | Remarks | Time (UT) | Activity | Remarks |
| 170746 | Start of SXR flare | AR9906 S14W34 | 210043 | Start of SXR flare | AR9906 S14W84 |
| 170750 | CME Start | | 210112 | $CME_a$ Start | |
| 170803 | CME appears at C2 | | 210114 | $CME_b$ Start | |
| 170824 | SXR flare Maximum | Class M2.6 GOES 08 | 210127 | $CME_a$ Reaches 3.3 $R_{SUN}$ in C2 | LASCO C2 V=2409 Km/sec |
| 170826 | CME Reaches 3.7 $R_{SUN}$ in C2 | LASCO V=1218 Km/sec | 210151 | SXR flare Maximum | Class X1.5 GOES 08 |
| 170840 | Type II/IV Onset | ARTEMIS-IV (cf. Figure 1) | 210127 | $CME_b$ Reaches 3.6 $R_{SUN}$ in C2 | LASCO C2 V=1264 Km/sec |
| 170957 | End of SXR Flare | | 210218 | $CME_a$ Reaches 13.4 $R_{SUN}$ in C3 | |
| 171000 | 2.5 % Decrease in Athens NM | From CME at 150350 | 210238 | End of SXR Flare | |
| 171027 | End of type II/IV | | 210318 | $CME_b$ Reaches 12.9 $R_{SUN}$ in C3 | |
| 171038 | CME Reaches 17.6 $R_{SUN}$ in C3 | LASCO | 211800 | Gradual Depression up to 2.5% in the Athens NM | From $CME_a$ |
| 171000 | 2.5 % Decrease in Athens NM | From CME at 150350 | 220800 | Gradual Depression up to 2.0% in the Athens NM | From $CME_b$ |
| 180800 | 3.0 % Decrease in Athens NM | From this CME | | | |

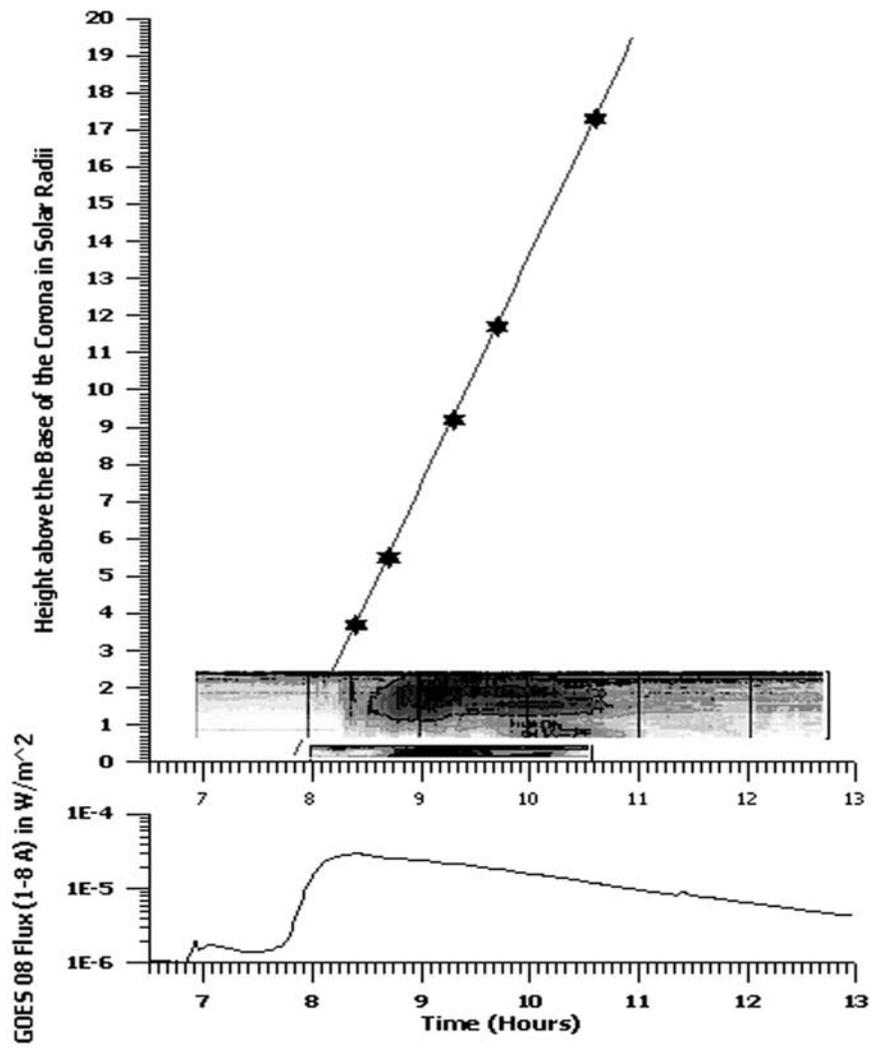

**Figure 1: Event of the 17 April 2002. Lower Panel: GOES 08 SXR flux from the 1-8 A channel. Upper panel: Time-Height plot of the associated CME; embedded are the ARTEMIS IV (lower inlay) and the Nançay Decametric Array (upper inlay) Dynamic Spectra**

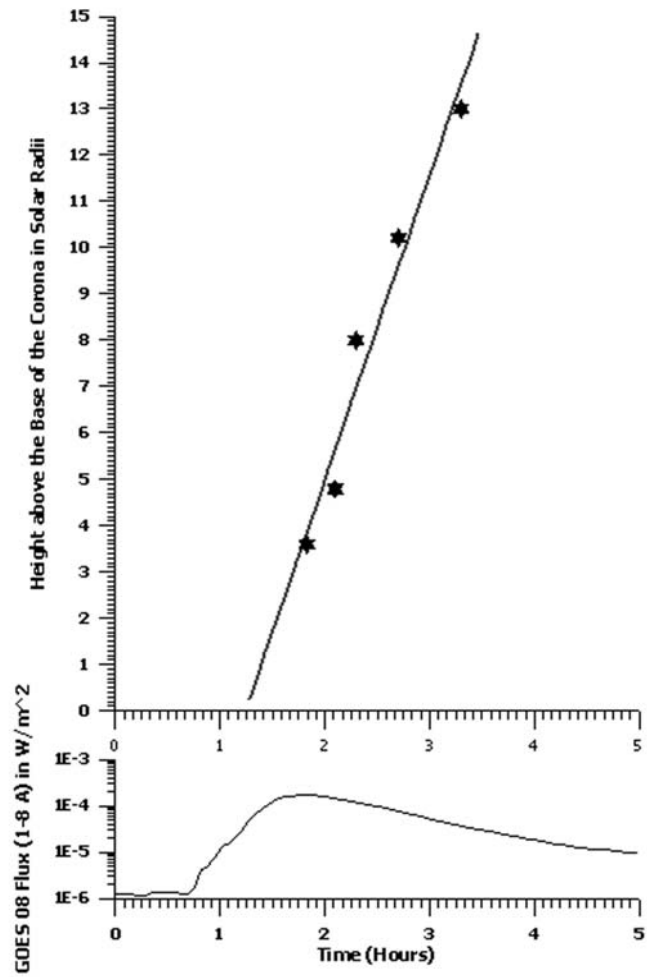

**Figure 2: Same as 1 but for the event of the 21 April 2002**

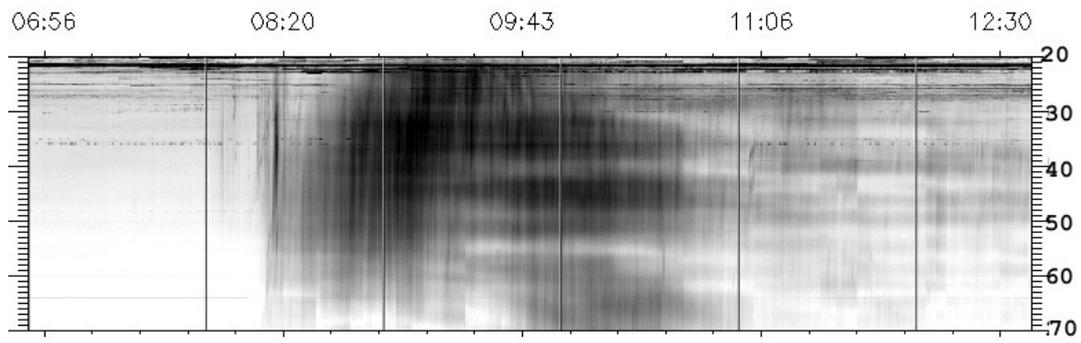

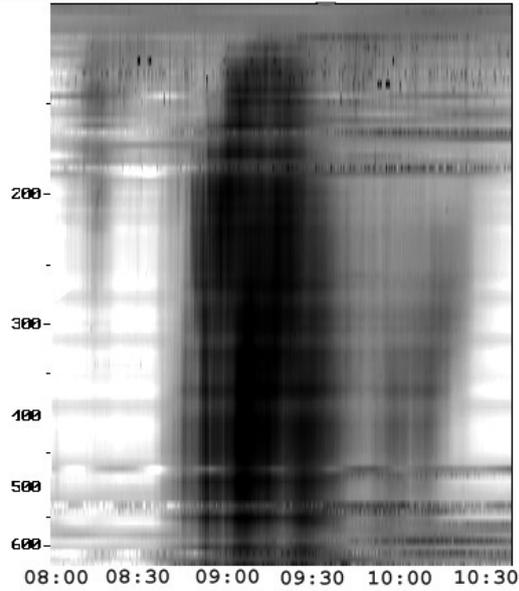

**Figure 3: Composite dynamic spectrum of the 17 April 2002 type II/IV radio burst. Upper Panel: Dynamic Spectrum in the 70-20 MHz range from the Nançay Decametric Array. Lower Panel: Dynamic Spectrum in the 110-698 MHz range from ARTEMIS IV**

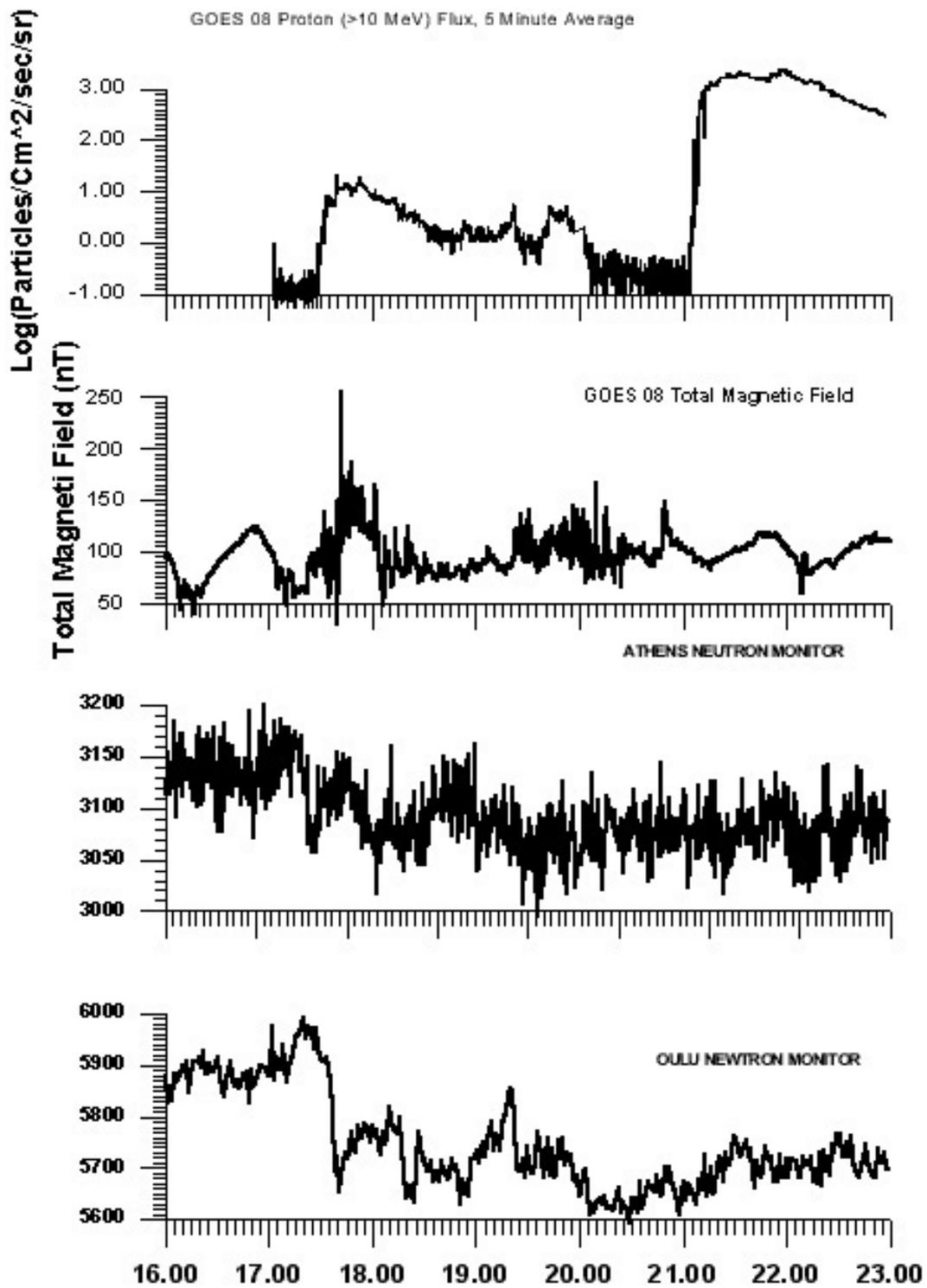

**Figure 4: Effects on the near Earth environment; top to bottom: GOES 08 proton flux (Energy > 10 MeV); GOES 08 total magnetic field; Athens and Oulu neutron monitor recordings**

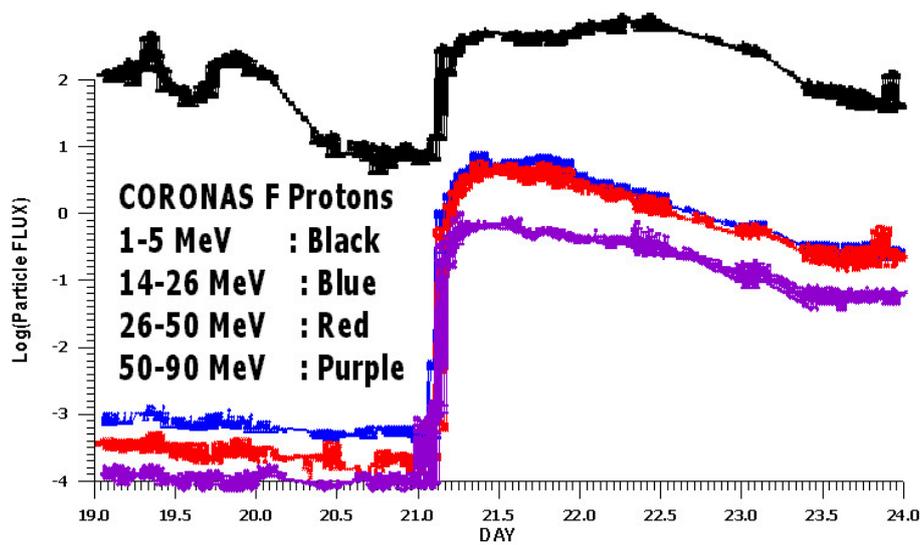

**Figure 5: CORONAS F proton flux (Energy bands: 1-5, 14-26, 26-50, 50-90 MeV) in the 19-24 April 2002 period.**